\documentclass[sigconf]{acmart}

\usepackage{subcaption}

\AtBeginDocument{%
  \providecommand\BibTeX{{%
    \normalfont B\kern-0.5em{\scshape i\kern-0.25em b}\kern-0.8em\TeX}}}

\acmPrice{15.00}

\copyrightyear{2023}
\acmYear{2023}
\setcopyright{rightsretained}
\acmConference[CHI '23]{Proceedings of the 2023 CHI Conference on Human Factors in Computing Systems}{April 23--28, 2023}{Hamburg, Germany}
\acmBooktitle{Proceedings of the 2023 CHI Conference on Human Factors in Computing Systems (CHI '23), April 23--28, 2023, Hamburg, Germany}
\acmDOI{10.1145/3544548.3581282}
\acmISBN{978-1-4503-9421-5/23/04}

\begin{document}





\title{Comparing Zealous and Restrained AI Recommendations in a Real-World Human-AI Collaboration Task}

\author{Chengyuan Xu}
\orcid{0000-0001-9981-8889}
\affiliation{%
  \institution{University of California, Santa Barbara}
  \city{Santa Barbara}
  \country{USA}
}
\email{cxu@ucsb.edu}

\author{Kuo-Chin Lien}
\affiliation{%
  \institution{Appen}
  \city{Sunnyvale}
  \country{USA}}
\email{klien@appen.com}

\author{Tobias Höllerer}
\affiliation{%
  \institution{University of California, Santa Barbara}
  \city{Santa Barbara}
  \country{USA}}
\email{holl@cs.ucsb.edu}

\renewcommand{\shortauthors}{Xu, Lien, and Höllerer.}

\begin{abstract}

When designing an AI-assisted decision-making system, there is often a tradeoff between precision and recall in the AI's recommendations. We argue that careful exploitation of this tradeoff can harness the complementary strengths in the human-AI collaboration to significantly improve team performance. We investigate a real-world video anonymization task for which recall is paramount and more costly to improve. We analyze the performance of 78 professional annotators working with a) no AI assistance, b) a high-precision "restrained" AI, and c) a high-recall "zealous" AI in over 3,466 person-hours of annotation work. In comparison, the zealous AI helps human teammates achieve significantly shorter task completion time and higher recall. In a follow-up study, we remove AI assistance for everyone and find negative training effects on annotators trained with the restrained AI. These findings and our analysis point to important implications for the design of AI assistance in recall-demanding scenarios.

\end{abstract}



\begin{teaserfigure}
  \includegraphics[width=\textwidth]{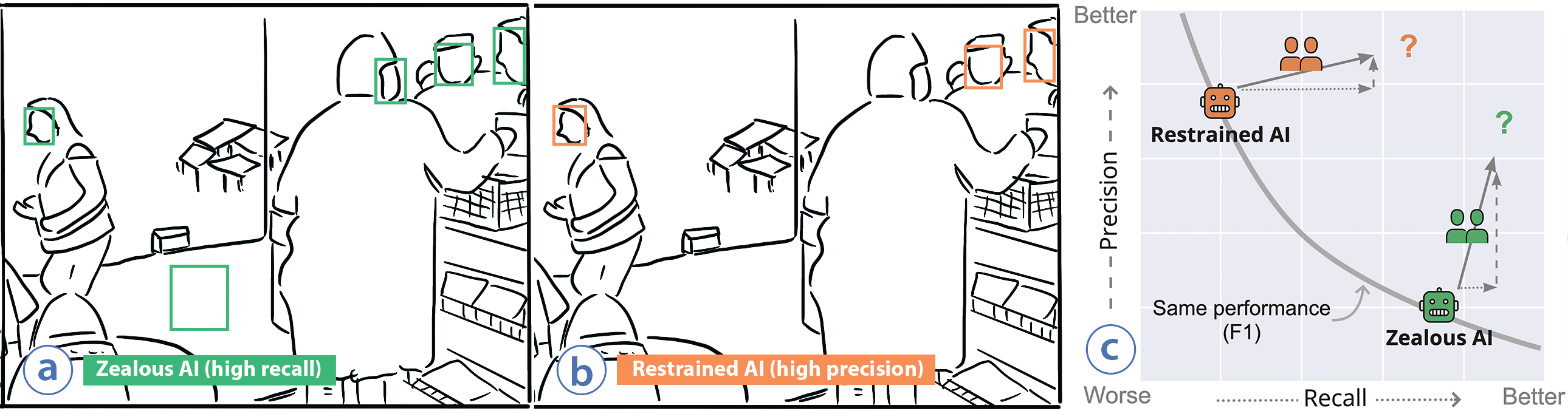}
  \caption{In video anonymization, face annotation and blurring is a high-stakes task that requires humans to check every frame. It demands high recall because one missed face can reveal a person's identity in the entire video. We can improve recall and reduce task completion time by forming a human-AI team. We may have two AIs with the same (F1) performance as shown in (c) but provide different sets of recommendations~(a~\&~b). A "zealous"~AI would prioritize recall by suggesting more detections, even low-confidence ones. A "restrained"~AI would only provide high-precision recommendations. Which AI teammate can help the human annotators finish in less time and with higher recall?}
  \Description{A teaser figure that shows two different methods for face annotation used for comparison in this paper. The left image shows an AI that tends to over annotate, the middle image shows an AI that tends to under annotate, on the right side a precision-recall figure shows both AI in fact have similar performance in F1 score.}
  \label{fig:teaser}
\end{teaserfigure}

\maketitle

\section{Introduction}\label{intro}

Machine-learning-based artificial intelligence (AI) systems have exceeded human performance in certain applications. But in high-stakes domains where fully-autonomous AI is not at peak performance or not permitted, such as in clinical decision-making \cite{bussone2015explanation, caruana2015intelligible, wang2016deep, xie2020chexplain, van2022fp} or driver assistance \cite{10.1145/3411764.3445351, 8798999, Gopinath_2022_CVPR}, forming a human-AI team is a viable strategy to improve both efficiency and accuracy. AI can provide recommendations while human users maintain agency and control over the final decisions. Studies have shown the human-AI team is expected to achieve "complementary team performance" -- the team performance being better than either one alone \cite{bansal2021accurate, Bansal_2021, Zhang_Complete_Me}. But there are more questions than answers on which exact factors in the AI system affect the team performance and how.

Bansal et al. recently showed in simplified binary classification problems that \textbf{the most accurate AI is not necessarily the best teammate unless it helps to improve the team utility} \cite{bansal2021accurate}. But how about in more complex problems where the AI teammate is not simply better or worse for its accuracy? For example, in many computer vision problems, people determine the best-performing algorithms based on combination metrics such as the F1 score \cite{sasaki2007truth, csurka_what_2013}, which can be broken down into two metrics -- precision and recall \cite{davis_relationship_2006, lipton_optimal_2014, lin_microsoft_2014}. Researchers can either balance the two metrics or prioritize one over the other to identify the best model for their application \cite{erenel_improving_2013, morstatter_balance_2016}. Two AI systems can have the same F1 score but provide very different recommendations with different measures of recall (see a, b in Figure~\ref{fig:teaser}). The tradeoff between precision and recall puts them on different parts of the same F1 isoline (see Figure~\ref{fig:teaser}~c). Without additional context, one might argue that there is no better or worse between these two AIs.


In order to capitalize on complementary strengths of humans and AI when presented with tradeoffs in AI precision and recall, we need to be able to answer two questions: \textbf{1) for a given task, can we clearly identify if either precision or recall is more important than the other}, and \textbf{2)~independent of importance, is it vastly easier or harder for humans to improve either precision or recall.}

Consider for example a pedestrian detection task in a driver assistance system: prioritizing the detection model towards either precision or recall will hurt the other. Human instinct tells us the risk of a missing detection could be lethal, so we should tune the AI system to prioritize recall, i.e., towards a "zealous" AI that provides more detections (recommendations), even the low-confidence ones, at the risk of more false positive errors. In this context, the opposite "restrained" AI would only provide high-confidence detections and prioritize precision, but at the risk of more false negative errors.


In this work, we investigate how a high-recall zealous AI and a high-precision restrained AI can affect human-AI team performance in a real-world scenario. Compared to, say testing pedestrian detectors on the road, video anonymization is a similar but easier-to-test recall-demanding task. We set up a face annotation task for personally identifiable information~(PII) protection that blurs human faces in a real-world video dataset \cite{Ego4D2022CVPR}. PII protection is a critical task with increasing demand for both ethical research and abiding by regulatory requirements\footnote{E.g., The General Data Protection Regulation (EU) or The California Consumer Privacy Act of 2018 (CCPA)}. Similar to pedestrian detection, where the cost of a missing detection is very high, one unlabeled face in a single frame can reveal a person's identity in the entire video, if not the entire dataset. 

This paper focuses on the common yet critical human-AI collaboration setting, in which recall is more important than precision. As for our second question, "is it vastly easier or harder for humans to improve either precision or recall?", an in-depth analysis of the video annotation workflow shows that improving recall is more costly than precision in this task since it is much harder for human annotators to draw a bounding box accurately than rejecting an incorrect one (see Section \ref{sec:pilot} \& \ref{sec:FPR} for a full discussion).

The answers to our two questions for our task reveal \textbf{an optimization opportunity: the AI recommendation tradeoff between precision and recall can be used to exploit complementary strengths of the human and the AI in such collaborative tasks}. We posit that similar optimization opportunities exist for many other human-AI collaboration tasks. In addition, locating faces is a human instinct\footnote{Here we refer to the ability to find human faces in a given image. We do not refer to recognizing people by face, which can be affected by Prosopagnosia (face blindness).} that requires no specific training or domain expertise to get started, making face detection a good candidate task to study the effects of different AI recommendations. The relatively small inter-personal differences also make the task a good representative of recall-demanding human-AI collaboration tasks. 

\begin{figure}[t!]
  \centering
  \includegraphics[width=1.\linewidth]{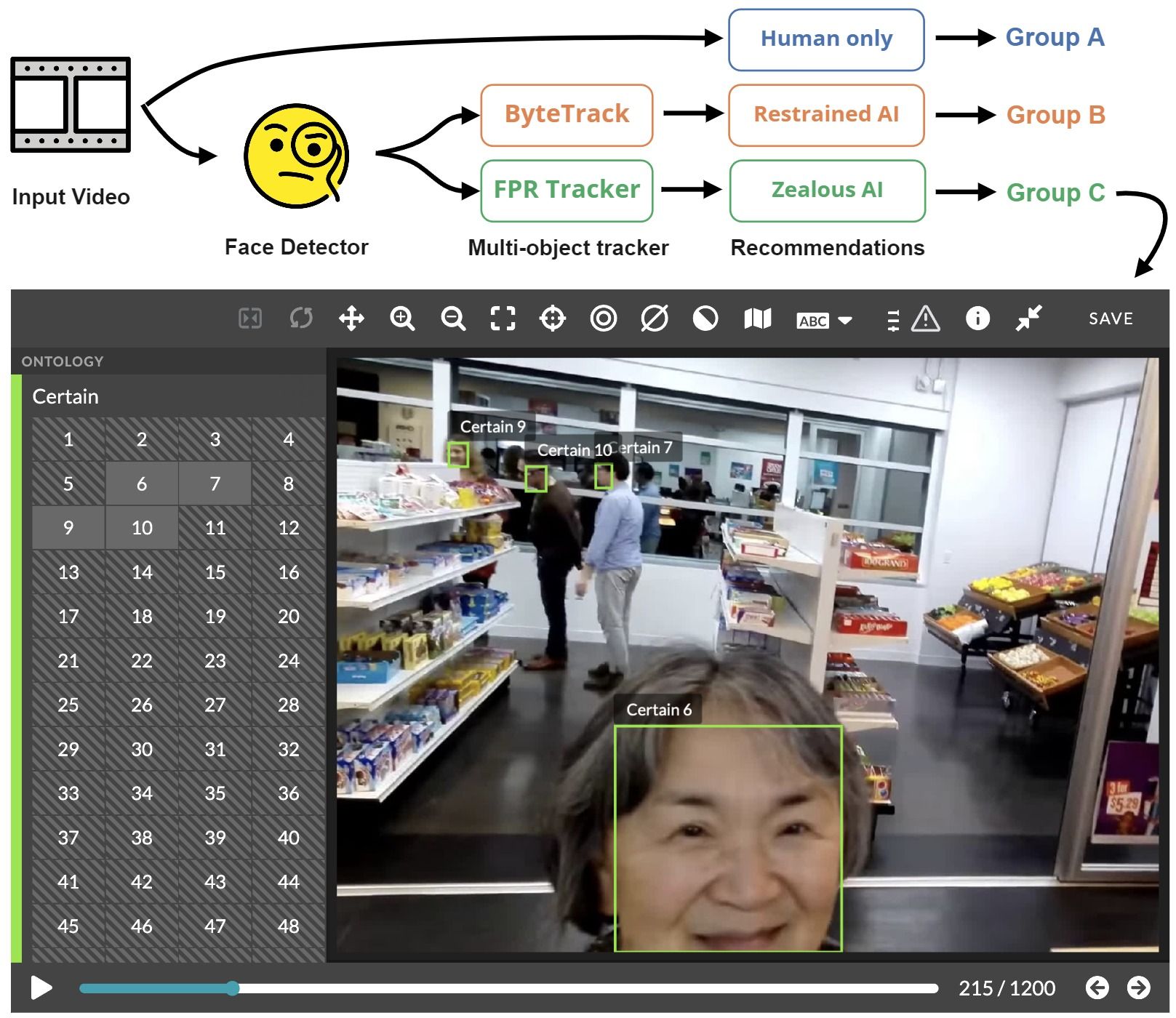}
  \caption{Data processing workflow for Part 1 of the study and the annotation tool user interface. The two AI teammates share the same face detector, which generates bounding box face detections for each frame independently. The ByteTrack tracker \cite{zhang2022bytetrack} and our proposed false-positive-robust~(FPR) tracker define the restrained or zealous AI recommendations -- they track the per-frame detections temporally to pre-annotate the videos as shown above. For the human-only workflow, annotators must manually draw a box and adjust its size and location across many frames.}
  \Description{A flow chart image that shows how the three treatment groups receive the video annotation and a screen shot of the annotation tool interface.}
  \label{fig:tool_interface}
\end{figure}

Our large-scale empirical study had 78 professional data annotators spend over 3,466 person-hours\footnote{Our system logged 3,466 person-hours of annotation work, which does not include pilot studies, training sessions, and answering multiple questionnaires. On average it took the 78 annotators three to four weeks to finish the entire study.} to submit a total of 2780 annotated 30-second videos. The between-subjects study split the annotators into three 26-people treatment groups. Detailed worker profiles ensured similar average experience between the groups (details in Section~\ref{sec:participants}). Each participant annotated human faces in 36 real-world videos of a variety of activities (see examples in Figure~\ref{fig:ego4d_examples}). We measure each group's annotation quality and task completion time. Any improvement in time is very meaningful for annotation tasks not only because of the cost. Fatigue induced by long working hours may also cause a decline in quality. 



In Part 1 of the \textbf{two-part study}, the three groups of annotators processed the same 24 videos, each with a) no AI assistance, b) pre-annotated bounding boxes recommended by the restrained AI, or c) the zealous AI. Figure~\ref{fig:tool_interface} summarizes the treatment groups and shows the annotation tool's interface. In Part 2, the three groups annotated another 12 videos but all without the AI's help. This design \textbf{allows us to learn how prior human-AI collaboration experience can affect user skills}, should they lose access to AI recommendations in the future. The two-part experiment aims to answer the following research questions:
\begin{itemize}
  \item[Q1] Can the human-AI teams achieve "complementary team performance" in this task?
  \item[Q2] Which AI helps annotators be more efficient, i.e. save time?
  \item[Q3] Which AI helps annotators achieve higher recall?
  \item[Q4] Will collaborating with an AI improve or hurt user skills?
\end{itemize}


We will answer each of the research questions in Section~\ref{sec:results}. Here we summarize this work's contributions:
\begin{itemize}
  \item We propose the concept of restrained and zealous AI recommendations to compare the tradeoff between precision and recall in tuning AI-assisted decision-making systems and investigate how they affect human-AI team performance in high-stakes recall-demanding tasks. 
  \item We design a large empirical study to compare the restrained and zealous AI on a face annotation task for video anonymization with 78 professional data annotators. The two-part experiment yielded significant findings to inform future AI assistance design for recall-demanding tasks.
  \item The analysis of 3,466 person-hours of annotation work reveals significant findings:
    \begin{itemize}
        \item Our study serves as a real-world case study of complementary team performance (cf.  \cite{nagar2011making,Kamar2012Combining,patel2019human, lai_22}).
        \item Identifying the complementary strengths of both human and AI teammates for a task is key to better team performance. The recall-demanding task and the higher cost of improving recall motivated us to propose the zealous AI, which provides high-recall recommendations and leads to significantly better task completion time and recall.
        \item The follow-up study demonstrates that naively pairing humans with an AI system designed for autonomous settings without optimizing it for the task at hand or for the human-AI workflow could potentially have a negative training effect on the users.

    \end{itemize}
\end{itemize}


\section{Related work}\label{related-work}

\subsubsection*{\textbf{Factors affecting human-AI team performance.}} While human-AI teams have been studied extensively from various perspectives like in crowdsourcing settings \cite{Kamar2012Combining, lundgard_2018}, computer vision tasks \cite{Kamar2012Combining, wang2016deep, patel2019human}, high-stakes tasks \cite{patel2019human, bansal2019beyond, bansal2019updates, zhang2020Effect}, and real-world tasks \cite{Kamar2012Combining, nagar2011making, patel2019human, amershi_2019, Wilder_2020, lai_22, van2022fp}, we still have more questions than answers on exactly which factors affect team performance and how. Researchers have looked into factors like users' mental models \cite{chakraborti2018algorithms, bansal2019beyond}, user expectations \cite{kay_2015, yin2019accuracyTrust, zhang_21}, cognitive biases \cite{rastogi2022cognitive}, model updates during collaboration \cite{bansal2019updates}, model accuracy \cite{yin2019accuracyTrust, bansal2021accurate}, model interpretability or explanations \cite{bilgic2005explaining, ribeiro2016why, lundberg2017unified, Feng19Interpret, wang2019designing, kaur2020interpreting, Bansal_2021}, as well as the tradeoff between accuracy and interpretability \cite{caruana2015intelligible}. Studying user's trust and appropriate or inappropriate reliance on AI \cite{muir1987trust, bussone2015explanation, yu2017trust, kunkel2019explain, lu2021reliance, zhang2020Effect} is another important direction.

This paper is aligned with works that focused on the tradeoff between precision and recall in AI recommendations and its effect on team performance. Kay et al. \cite{kay_2015} introduced the acceptability of accuracy as a new measure and survey instrument to connect classifier evaluation to users’ subjective perception of accuracy. Kocielnik et al. \cite{kocielnik2019will} compared two 50\%-accurate AI-powered scheduling assistants -- one avoids false positive errors, and one avoids false negative. This is a similar design as for our restrained and zealous AIs -- their study found that false positive errors are more acceptable by participants, which corroborates the overall better performance we observed in the zealous AI group, who also dealt with more false positive errors.

Balancing precision and recall to compare two real-world AI systems in a human-AI collaboration task is not easy, previous works derived insight from hypothetical systems or manually balanced recommendations \cite{kay_2015, kocielnik2019will}. In this work, we provide a real-world user study by observing how 78 professional users would interact with two high-performance face tracking AI systems that are tuned to truthfully portray the realistic tradeoff between high-precision and high-recall on a recent egocentric video dataset.


\subsubsection*{\textbf{Face detection.}} The annotation platform we used has a built-in face detector, RetinaFace \cite{DBLP:journals/corr/abs-1905-00641}, integrated for autonomous workflows. Our literature search found RetinaFace remains a top-ranking method on the WIDER FACE benchmark \cite{yang2016wider}. Because more recent methods do not provide significant performance improvement, we continue to use RetinaFace as a consistent baseline to compare with our algorithmic improvements in tracking. 

\subsubsection*{\textbf{Multi-object tracking.}} In the AI-assisted face annotation task, the AI teammate provides annotation recommendations for users to review. Conventionally a face detector provides per-frame face bounding boxes and a multi-object tracking (MOT) algorithm produces continuous tracks of the same object across frames. This is known as tracking-by-detection. Recent MOT methods like TransTrack \cite{sun2020transtrack}, DETR \cite{10.1007/978-3-030-58452-8_13}, Deformable DETR \cite{zhu2020deformable}, TrackFormer \cite{meinhardt2021trackformer}, and TransMOT \cite{chu2021transmot} etc. all move toward the end-to-end Transformer-based \cite{NIPS2017_3f5ee243} architecture. However, these black-box MOTs share the same drawback as they are designed for fully-autonomous settings. Similar to Caruana et al.'s observation that modular system provides better transparency \cite{caruana2015intelligible}, the two-part tracking-by-detection frameworks actually provide us the interpretability and flexibility to steer the output recommendations as needed, so we can produce restrained and zealous AI recommendations for comparison. We reviewed state-of-the-art methods in related multi-object tracking benchmarks \cite{MOT16,MOTS20,Dave:2020:ECCV} in search of a multi-object tracker suitable for a human-in-the-loop annotation workflow.  ByteTrack \cite{zhang2022bytetrack} is a conventional tracker that outperforms numerous Transformer-based trackers mentioned earlier.

\subsubsection*{\textbf{Video annotation.}} While there are various public video annotation platforms or tools to choose from \cite{10.1145/3343031.3350535, DBLP:journals/ijcv/VondrickPR13, 5459289, DBLP:journals/mta/KavasidisPSGS14}, we use a proprietary video annotation tool to gain access to professional data annotators who are already familiar with the specific tool from their past project experience. This tool has Linear Interpolation~\cite{DBLP:journals/ijcv/VondrickPR13} activated by default, which provides semi-automatic assistance by linearly interpolating a box between two manually annotated key frames. In this study, all participants, including annotators who review AI's annotation recommendations have access to this functionality. Linear Interpolation is also an ideal baseline as all participants have sufficient experience using it. We will refer to this basic setup as human only, the baseline method, or the manual method in the rest of the paper.

\section{Algorithm choices and pilot studies}

\subsection{Precision and recall in multi-object tracking}

Precision, recall, and F1 are important performance metrics that can describe the characteristics of a model and are central concepts in this work and other human-AI research \cite{kay_2015, kocielnik2019will}. Specifically, in the context of annotating and tracking faces with bounding boxes in videos:
{\small
\begin{align}
    \label{eq:precision}
    \text{Precision} &  = \frac{\text{TP}}{\text{TP + FP}} = \frac{\text{Face boxes correctly drawn}}{\text{All boxes drawn by the user (or the AI)}} \\
    \label{eq:recall}
    \text{Recall} & = \frac{\text{TP}}{\text{TP + FN}} = \frac{\text{Face boxes correctly drawn}}{\text{All ground truth face boxes}}   \\
    \label{eq:f1}
    \text{F1} & = \frac{2\cdot\text{precision}\cdot\text{recall}}{\text{precision}+\text{recall}}
\end{align}
}where the TPs are true positives, face boxes that were correctly drawn. The FPs are false positives, boxes drawn by the AI or user which did not match real faces properly. The FNs are false negatives, where there is a real face, but the box is missing.

The F1 score is the harmonic mean of the precision and recall (Equation~\ref{eq:f1}). We visually introduced the concept of this function using three methods that have the same F1 score in Figure~\ref{fig:teaser} (c). Put simply, a video pre-annotated by a high-recall method (zealous AI) would have more false-positive boxes -- the user will make more rejections but add fewer missing boxes. A video pre-annotated by the high-precision method (restrained AI) would provide mostly correct boxes but the user will need to add more missing boxes.

We are interested in how users will perform differently given restrained or zealous AI recommendations in an AI-assisted face annotation task. While it is easy to generate high-precision annotations by simply avoiding low-confidence detections, it is hard for trackers to produce high-recall results while maintaining a similarly high F1 score at the same time. This motivates us to propose a tracking algorithm that pushes recall to the limit, but aims to maintain a similar level of F1 score. We take advantage of the fact that \textbf{our tracking results will be reviewed by human annotators, allowing us to make targeted optimizations}. We test our ideas of a user-friendly tracker with professional annotators through pilot studies. Observing how users work with trackers allows us to further improve the algorithm.

\newpage
\subsection{Pilot studies}\label{sec:pilot}

We conducted two pilot studies to observe how professional data annotators work with AI recommendations. Annotators were tasked to draw bounding boxes around potentially moving or blurred faces of any size in a 1,200-frame video sequence of a busy shopping scene in both sessions (similar to hard videos in the formal study). We provided training material on how to review recommendations from the AI for the face annotation job. The annotation tool user interface is shown in Figure~\ref{fig:tool_interface}. With their consent, we recorded their screens to keep track of mouse movements and other user habits. Each session included ten different users with above-average experience. Both pilot studies concluded with a survey about experiment design and their experience. The two pilot sessions were spaced two weeks apart to test algorithm and design improvements.

Users' screen recordings helped us observe the following user habits and behaviors that are not possible to be identified solely from the results:
\begin{itemize}
  \item \textit{Certain bad recommendations cost most of the human review efforts.} Following the Pareto Principle \cite{juran2005quality}, annotators in fact spent most of their time and effort amending a small fraction of AI recommendations. The tiny bounding boxes (see examples of three tiny faces in Figure~\ref{fig:tool_interface}), duplicate detections (often clustered), and temporally sparse detections (short tracks) are the most costly recommendations. Addressing these issues allows annotators to have better continuity in their workflow.
  \item \textit{Model explanation should not increase task complexity.} Initially, we offered model explanations using "Certain" and "Uncertain" labels based on the face detector's confidence, hoping this can assist users' decision-making. But video recordings and user feedback revealed that the extra information in fact increased the task complexity and caused unnecessary confusion. This design was eventually not considered in the formal experiment.
\end{itemize}

\begin{figure*}[t!]
  \centering
  \includegraphics[width=1.0\linewidth]{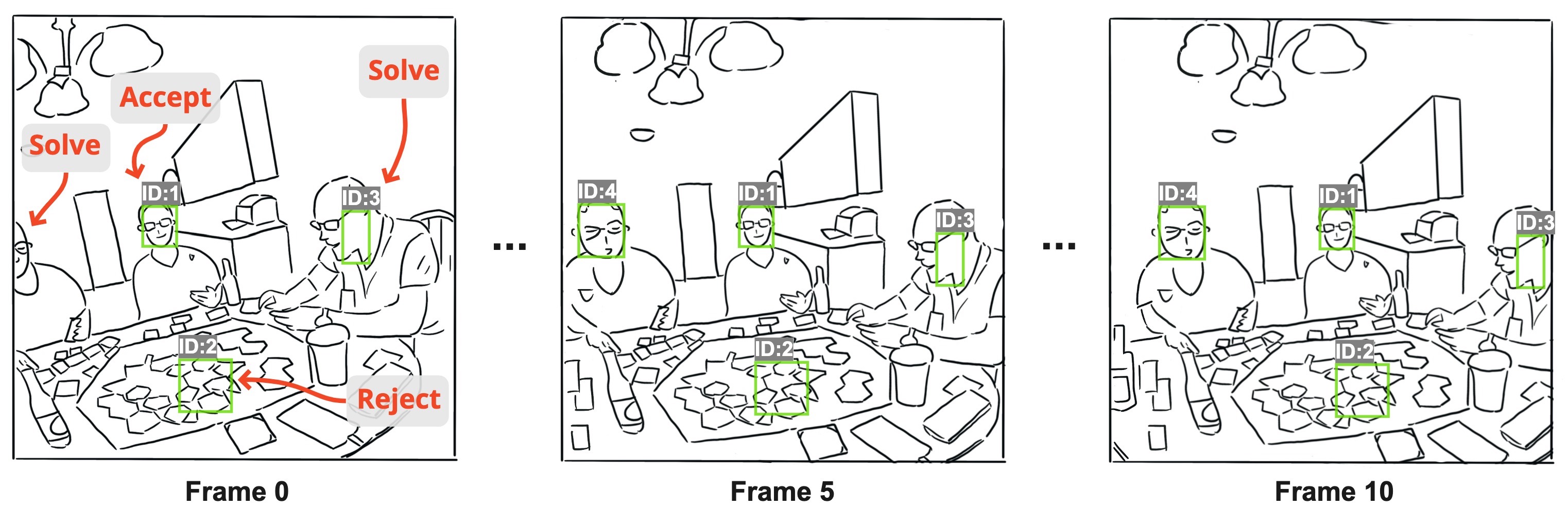}
  \caption{When reviewing the AI teammate's recommendations (green bounding boxes), a user takes one of the three actions for each box: \textit{\textbf{accept}}, \textit{\textbf{reject}}, or \textit{\textbf{solve}}. In video annotation, because \textbf{the boxes are temporally tracked across many frames, each action's time complexity is drastically different}, note the two types of \textit{Solve} in frame 0 can come at different cost, too. \\ \hspace*{1em} ID:1 -- A user can \textit{accept} the true-positive track ID:1 boxes without any action. \\ \hspace*{1em} ID:2 -- The entire false-positive ID:2 track can be rejected with two mouse clicks by deleting the ID in any of the frames, which is $O(1)$ in time complexity. \\ \hspace*{1em} ID:3 -- False-positive recommendations, like track \textbf{ID:3}, are the most time-consuming to \textit{\textbf{solve}}: the user can delete and redo this face, or manually adjust every frame until the AI's pre-annotation becomes acceptable with $\geq n$ mouse clicks, $O(n)$. \\ \hspace*{1em} ID:4 -- In frame 0, to \textit{solve} the false-negative missing box for the left-most person, a user needs to manually draw a box and adjust its location and size until the AI-suggested box \textbf{ID:4} comes in with $\leq n$ mouse clicks, $O(n)$ where $n$ is the number of frames.}
  \Description{Three frames of hand-drawn images show the three actions of accept, reject, or solve on different types of AI-recommended boxes. A missing or incorrect face detection requires to solve, a false-positive detection requires to reject, and true-positive detections require no action.}
  \label{fig:user_actions}
\end{figure*}

Observing how human annotators review AI recommendations (bounding box pre-annotations) in multi-object detection and tracking tasks inspired us to \textbf{break the complex workflow into three fundamental user actions:} \textit{\textbf{accept}}, \textit{\textbf{reject}}, or \textit{\textbf{solve}}, each coming with a higher cost in time. Figure~\ref{fig:user_actions} explains each action's time complexity. We can connect these three actions with our two main objectives (time and recall) to make \textbf{a simple deduction to identify the human-AI complementary strengths} in this task:
\begin{itemize}
  \item[1] \textit{\textbf{reject}} improves precision and \textit{\textbf{solve}} improves recall. A correct \textit{\textbf{accept}} improves both.
  \item[2] It takes the AI constant time to \textit{\textbf{solve}} additional cases (give more recommendations) with a downside of more false-positive boxes for humans to reject.
  \item[3] Humans are faster at \textit{\textbf{reject}}ing a false-positive (incorrect) box than to \textit{\textbf{solve}} a false-negative (missing) box.
  \item[4] We also know recall is more important than precision in video anonymization tasks.
  \item[5] Thus, \textbf{a clear path to better human-AI team performance is to delegate more \textit{\textbf{solve}} actions to the AI, so the human's overall effort is reduced by doing more easy \textit{\textbf{rejecting}} and only \textit{\textbf{solving}} the most challenging faces}.
\end{itemize}

\newpage
\subsection{The false-positive-robust (FPR) tracker}
\label{sec:FPR}

We adopted a tracking-by-detection system to produce face pre-annotations (Section~\ref{related-work}), the two-part system design allows us to feed the same per-frame face detection from RetinaFace \cite{DBLP:journals/corr/abs-1905-00641} to different downstream multi-object trackers like the ByteTrack  \cite{zhang2022bytetrack} or our own designs for a fair comparison. Learning from our pilot studies observations, we propose the false-positive-robust (FPR) tracker that specifically provides user-friendly annotation recommendations. We use \textbf{the following unconventional strategies} to design the FPR tracker that can take overwhelmingly noisy detections with a high false positive rate as input but outputs "clean" tracks for a human-in-the-loop workflow:
\begin{itemize}
  \item To improve the AI's recall, we apply an \textbf{extremely low threshold ($t\geq0.01,~t\in[0,1]$) on the face detector's confidence score} to keep any potentially useful detected boxes. This is not a viable solution for Autonomous AI systems but we are working in conjunction with a human. 
  \item The consequence of such a low face detector threshold is \textbf{clusters of overlapping boxes} on small faces. Our solution: for each cluster, we perform non-maximum suppression \cite{neubeck_non_max} by only keeping the single bounding box with the highest confidence score because in most cases they are duplicate detections on one true face. This step also improves the AI recommendations' precision.
  \item Finally, based on our observation that \textbf{the majority of temporally sparse detections are false positives} induced by the low threshold, we remove any tracks that are shorter than $m$ consecutive frames so they do not interrupt users' continuity. We used $m=10$ in the FPR tracker. Although some true-positive faces are also removed, users are much faster at solving an unlabeled face from scratch than filling the gaps between temporally sparse detections.
\end{itemize}

To design the experiment, we also need a restrained AI that generates recommendations of similar performance (F1~score) but with high precision. This is done by using only the high-confidence ($t\geq0.8, t\in[0,1]$) face detections with ByteTrack. To ensure fair comparison and reduce moving parts in our systems, we use the same face detection model RetinaFace \cite{DBLP:journals/corr/abs-1905-00641} for both AI teammates. It is the two different (fully transparent) trackers we apply that push the AI recommendations towards either high-precision or high-recall (Figure~\ref{fig:tool_interface}).

Note that we were only able to optimize the FPR tracker and ByteTrack through pilot studies because the ground truth data was not available for the 36 videos used in the user study. After the study, we aggregated the annotations from all 78 participants (2,780 submissions in total) to form an expert-reviewed consensus to serve as the ground truth. It turns out the zealous AI recommendations (FPR tracker) yielded an F1 score of $90.9\%$ and the restrained AI (ByteTrack) had an F1 score of $93.4\%$. While the two AIs did not provide identical initial performance for their human teammates, we achieved the goal of two distinctive high-recall and high-precision AIs (Figure~\ref{fig:team_effort}). The performance gap also provided us additional evidence to support our previous deduction on the zealous AI being the superior choice for this task, which we will discuss in Section~\ref{complementary performance}.

\begin{figure*}[t!]
  \centering
  \includegraphics[width=1.0\linewidth]{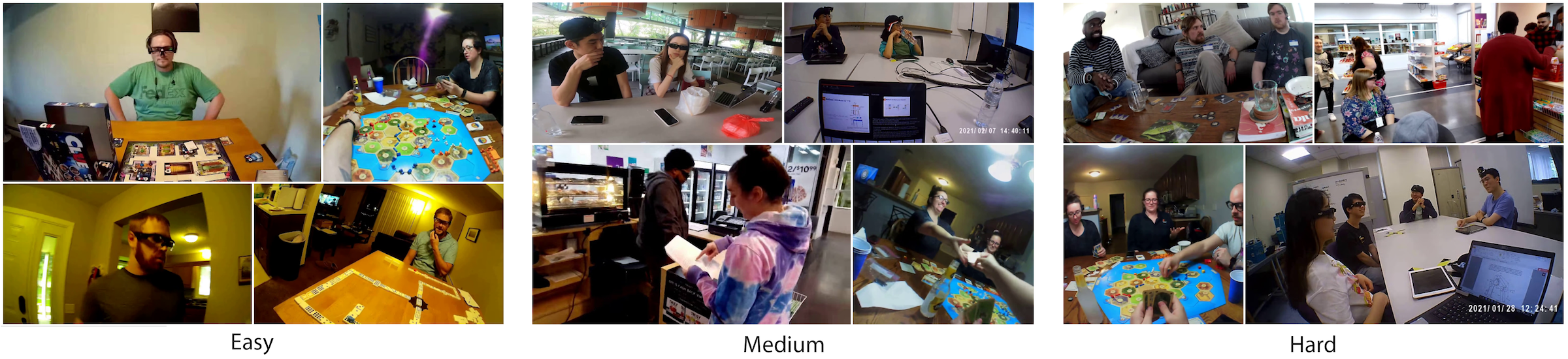}
  \caption{Screenshot examples of Ego4D videos \cite{Ego4D2022CVPR} used in our face annotation experiment. Easy videos include about one face to annotate in each non-empty frame. Medium videos include about two faces. Hard videos include three or more faces. Videos with more faces are expected to take longer time to finish. The study results show shorter to longer completion times for Easy, Medium, and Hard videos in both parts (see Figure~\ref{fig:part1_task_time} and Figure~\ref{fig:part2_task_time}), demonstrating that our video difficulty categorization is reasonable and performed as expected. We also considered scene diversity, box size (smaller faces are harder), and camera movement intensity (more movement is harder) to ensure a balanced difficulty distribution in selecting the specific videos.}
  \Description{Three clusters of images that show examples of videos in different difficulty levels.}
  \label{fig:ego4d_examples}
\end{figure*}

\section{Experiments}

In this work, we aim to investigate how restrained and zealous AI recommendations will affect human-AI team performance. We are also curious if the collaboration experience with an AI teammate can affect users' skills, should they lose access to AI assistance in the future. We design a two-part empirical study to test the restrained and zealous AIs in a recall-demanding high-stakes task.

\subsection{The task and data.\label{task_data}} Face annotation for video anonymization is a perfect example of recall-demanding tasks -- a missing face in a single frame can reveal a person's identity in the entire video. The high-stakes nature requires humans to annotate or verify every frame, yet the manual process will become the throughput bottleneck. The tedious process and long hours may also fatigue annotators and cause a decline in quality. In addition, \textbf{because the task of locating faces requires no specific training or domain expertise, it should help the generalizability of our observations} to other AI-assisted annotation tasks or even to other recall-demanding human-AI collaboration tasks.

In our human-AI collaboration setting, the AI teammate provides recommendations in the form of bounding boxes (see examples in Figure~\ref{fig:tool_interface}), and a user reviews each of the AI's pre-annotations to make one of the three decisions shown in Figure~\ref{fig:user_actions}. We evaluate users' performance on the two most important metrics for face anonymization: \textbf{task completion time} and \textbf{recall}.

To test different AI recommendations in a real-world setting, we curate 36 first-person videos from a large-scale egocentric video dataset Ego4D \cite{Ego4D2022CVPR}. Privacy has always been a major concern for datasets collecting human activities so first-person videos are ideal for this study. The videos we selected include various indoor social activities that are suitable for benchmarking face detection and annotation tasks. Each video clip is 30 seconds long, or 900 frames. We estimate each video takes about 30 minutes to one hour to fully annotate, depending on its difficulty.

The different annotation methods (without or with different AI recommendations) adopted by the three treatment groups are the first level of independent variables that we will discuss in the next section. The second level of independent variables that can affect users' performance is the difficulty of the videos. We divide the videos into Easy, Medium, and Hard categories based on the average number of people one needs to track simultaneously in non-empty frames (see examples in Figure~\ref{fig:ego4d_examples}). We also considered factors like scene diversity, bounding box size, and camera movement intensity that affect the annotation difficulty in a more subtle way. Based on this overall difficulty ranking distribution, we ensure Part 1 and Part 2 videos are not only similar in content but also consistent in annotation difficulty.

We generate the bounding box ground truth by aggregating the crowd's annotations to reach a consensus, which is further reviewed and refined by a domain expert. We used an equal number of manual and AI-assisted submissions for each video to generate an unbiased ground truth.

On task completion time, \textbf{annotators are advised to finish each video without taking breaks longer than five minutes} but we still need to reject outlier video completion times caused by a known limitation of the annotation tool -- the timer continues if an ongoing task window was left idle, or the timer will reset if the annotator continues from previously saved progress. We adopted median absolute deviation (MAD) \cite{LEYS2013764} by comparing each video's completion time within each group to reject 420 out of 2780 (15.11\%) completed videos, including completion times that are less than six minutes (the minimum time needed to verify each frame) or longer than median + 3 * MAD. The rejected videos also include all 36 submissions from one particular problematic user, see Section~\ref{sec:lower-bound}.

\begin{table*}[t!]
  \centering
  \begin{tabular}{ c|cc|cc|cc } 
  \toprule
    Group & Novice & Veteran & Part 1 method & Submissions & Part 2 method & Submissions \\
    \midrule
    A & 11 & 14 & Human only & 602 & Human only & 299 \\
    B & 14 & 12 & Restrained AI + Human & 619 & Human only & 304 \\
    C & 13 & 13 & Zealous AI + Human & 621 & Human only & 299 \\
    \bottomrule
  \end{tabular}
  \caption{In the two-part study, the three treatment groups use different methods in Part 1, but we remove all AI assistance in Part~2. The novice and veteran workers represent a balance of different user expertise in each group. The submission numbers are the 30-second annotated videos each group finished. Note that Group A is one user short as a particular worker was later rejected because of repeated bad submissions.}
  \label{table:treatments}
\vspace{-6mm} 
\end{table*}

\subsection{Participants and three treatment groups.\label{sec:participants}} A total of 78 in-house professional data annotators completed our study. It is important to note that in this project \textbf{they are paid at their regular hourly rate,  so participants are not motivated by compensation to work faster}.

In the between-subjects experiment, participants were evenly split into three 26-people treatment groups to annotate identical sets of videos. The annotators' profiles ensure similar average experience between the groups. The assignments also considered people's day/night shifts and computer setup to ensure a fair comparison.

The participants have at least two months or up to five years of data annotation experience, with an average experience of 20.9 months. We use the median experience of 17 months to split the user expertise factor so each group has about half novice and half veteran workers  (see Table~\ref{table:treatments}). All annotators were aware of participating in a study testing new AI-assisted annotation algorithms and were free to leave the study at any time. The Human Subjects Committee (HSC) approved our procedure and each participant was provided a consent form during the survey session.

Group A servers as the baseline, they use an efficient annotation tool that supports linear interpolation \cite{DBLP:journals/ijcv/VondrickPR13} but solely relies on manual annotation in both parts of the study. Groups B and C work with their AI teammates in Part 1 of the study. They use the same tool as Group A but the AI will have pre-annotated the videos (see example in Figure~\ref{fig:tool_interface}). Group~B reviews the restrained AI recommendations that prioritize precision. Group~C reviews the zealous AI recommendations that prioritize recall (see a, b in Figure~\ref{fig:teaser}). The treatment groups are summarized in Table~\ref{table:treatments} or Figure~\ref{fig:tool_interface}. We informed the participants in Groups B and C that they are working with an AI that provides recommendations to assist their annotation work, but they do not know the difference between the two human-AI groups.

\subsection{Experiment procedure of the two-part study.} Before beginning the study, we organized a video conference training session with each treatment group to calibrate the task background and requirements. All participants were also asked to review the instruction text and a training video on the landing page. Previous pilot study users become supervisors in each group to ensure all participants have finished the training and the surveys before processing to the next step. We also created three instant messaging (IM) groups to answer questions and send out reminders when necessary. The overall procedure can be summarized as follows:
\begin{align*}
  \text{Training} & \rightarrow \text{Survey 0} \rightarrow \\
  \text{Part 1 (24 videos, different methods)} & \rightarrow \text{Survey 1} \rightarrow \\ 
  \text{Part 2 (12 videos, same method)} & \rightarrow \text{Survey 2}
\end{align*}

In \textbf{Part 1}, all participants from Groups A, B, and C each annotated 24 videos using different methods. For each annotator, the videos were assigned in random order by the annotation platform. We also reminded all participants to avoid taking breaks longer than five minutes before finishing a video, so the timing is more accurate. Depending on the method and individual pace, it took all groups on the order of two to three weeks to finish Part 1. In \textbf{Part 2}, all participants annotated another 12 videos from similar scenes. But we took away the AI assistance from the two human-AI teams B and C in order to find out if their previous human-AI collaboration experiences trained them in any way so that they would perform differently on manual annotations from here on out. 

A post-task survey was administered after each part of the study. \textbf{Survey 0} was set to "repeat until perfect", this was to verify that the participants were clear about the task requirements before they could start the actual annotation. \textbf{Survey 1} focused on getting people's immediate feedback on their experience working with the AI they were paired with. Questions include the correctness and consistency of the AI recommendations, and if the AI made their job easier. This allows us to compare if participants' subjective feelings match the different AI recommendations' underlying personae (high-precision vs. high-recall). \textbf{Survey 2} focused on comparing the annotators' preference between AI-assisted and human-only methods after they had experienced both workflows on the same task.


\section{Results}
\label{sec:results}

In this section, we present our study results and analysis by answering each research question presented in Section~\ref{intro}. For statistical analysis, we ran one-way ANOVA or one-way Welch ANOVA tests, depending on the underlying assumptions being satisfied, followed by Pairwise Tukey-HSD or Games-Howell post-hoc tests, respectively. To examine interactions between factors, we conducted two-way ANOVAs followed by Pairwise Tukey-HSD or Bonferroni-corrected post-hoc tests. We adopted Type III sums of squares in ANOVA to address unbalanced data.

\begin{figure}[ht!]
  \centering
  \includegraphics[width=1.0\linewidth]{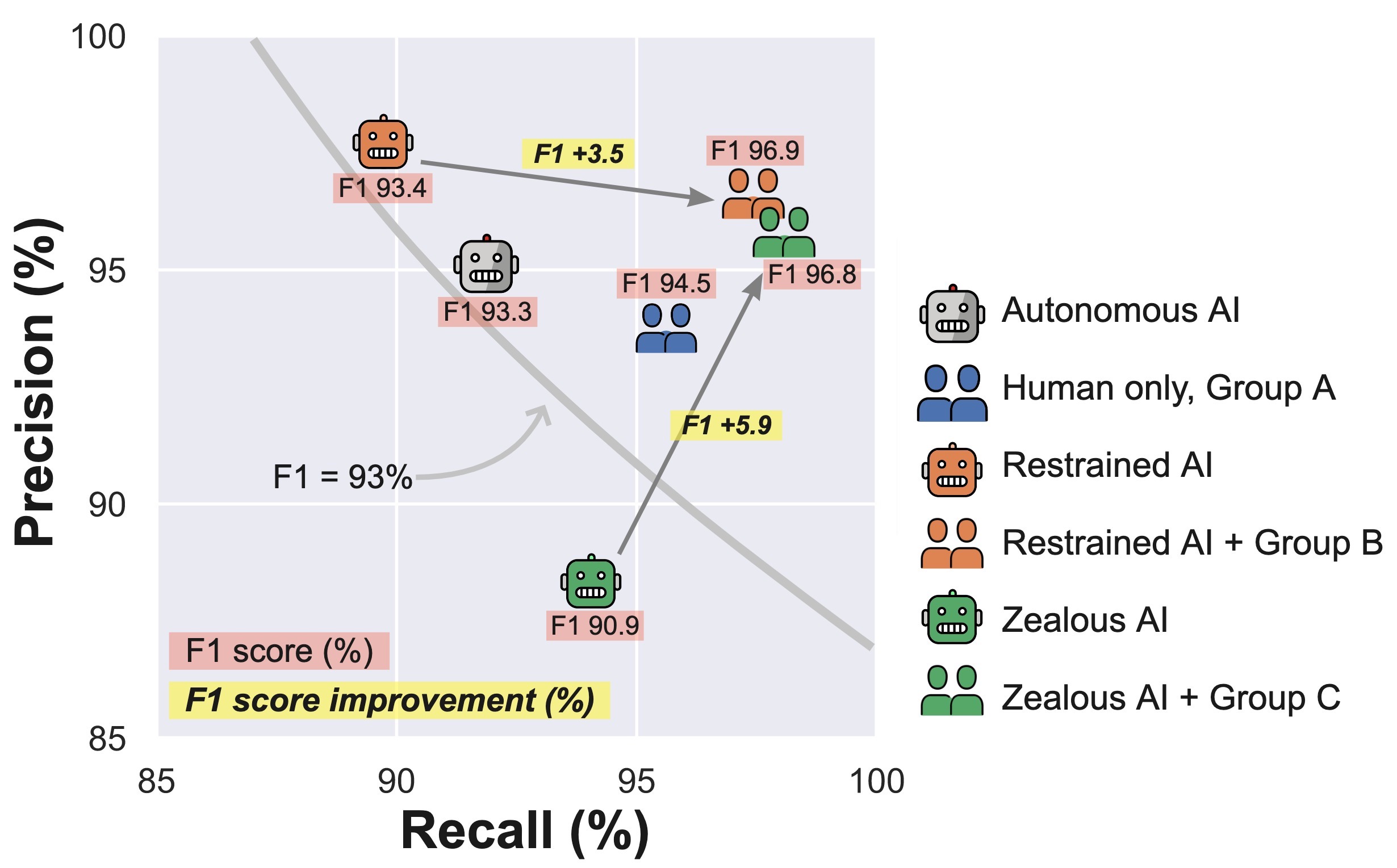}
  \caption{Visualizing each group's overall annotation quality on the precision-recall plot with F1 scores (Part 1). Group A manually annotates all videos and without surprise, they are the slowest (Figure~\ref{fig:part1_task_time}) with a quality better than Autonomous AI alone but worse than the two human-AI groups' team effort. Annotators in Groups B~\&~C had to \textit{\textbf{accept}}, \textit{\textbf{reject}}, or \textit{\textbf{solve}} the face boxes pre-annotated by the restrained or zealous AIs to improve the human-AI team performance. The arrows show how much humans improved from the AIs' initial annotation.}
  \Description{A figure plotting different treatment groups' annotation performance in precision and recall. The restrained AI started with high precision, and the zealous AI started with high recall.}
  \label{fig:team_effort}
\vspace{-6mm}
\end{figure}

\begin{figure*}
  \noindent \begin{minipage}[t]{0.48\textwidth}
    \includegraphics[width=\textwidth]{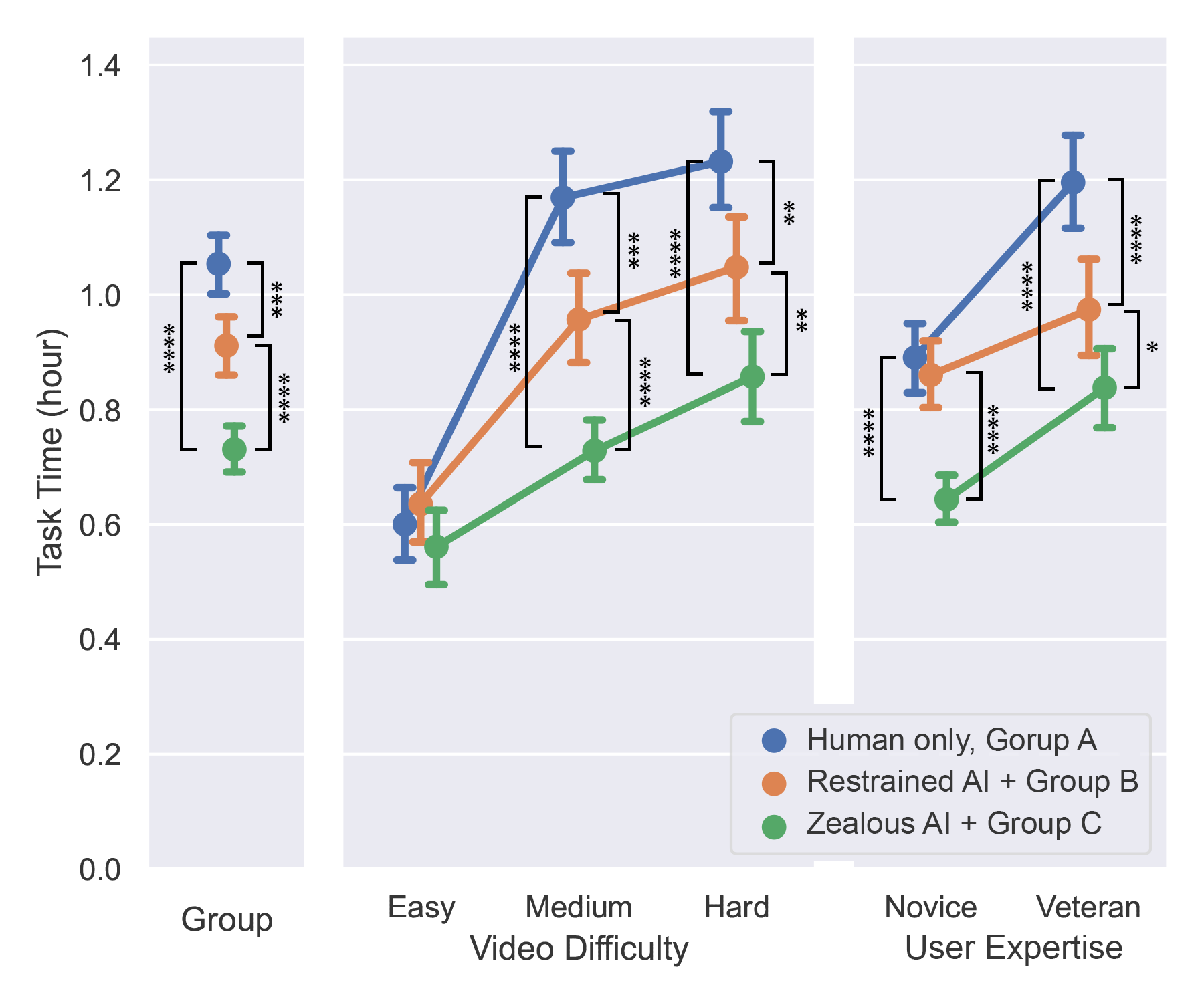}
    \vspace{-6mm}
    \captionof{figure}{Average annotation time for a single video in \textbf{Part~1}. Lower is better. Error bars represent the 95\% confidence interval. Treatment Group A used a baseline manual method and the annotators in Groups B and C reviewed restrained and zealous AI recommendations in Part 1. Groups B \& C included the GPU time used to calculate the AI recommendations.}
    \label{fig:part1_task_time}
    \Description{Part 1 task completion time of the three methods, details described in the main text.}
  \end{minipage}
  \hspace{0.02\textwidth}
  \begin{minipage}[t]{0.48\textwidth}
    \includegraphics[width=\textwidth]{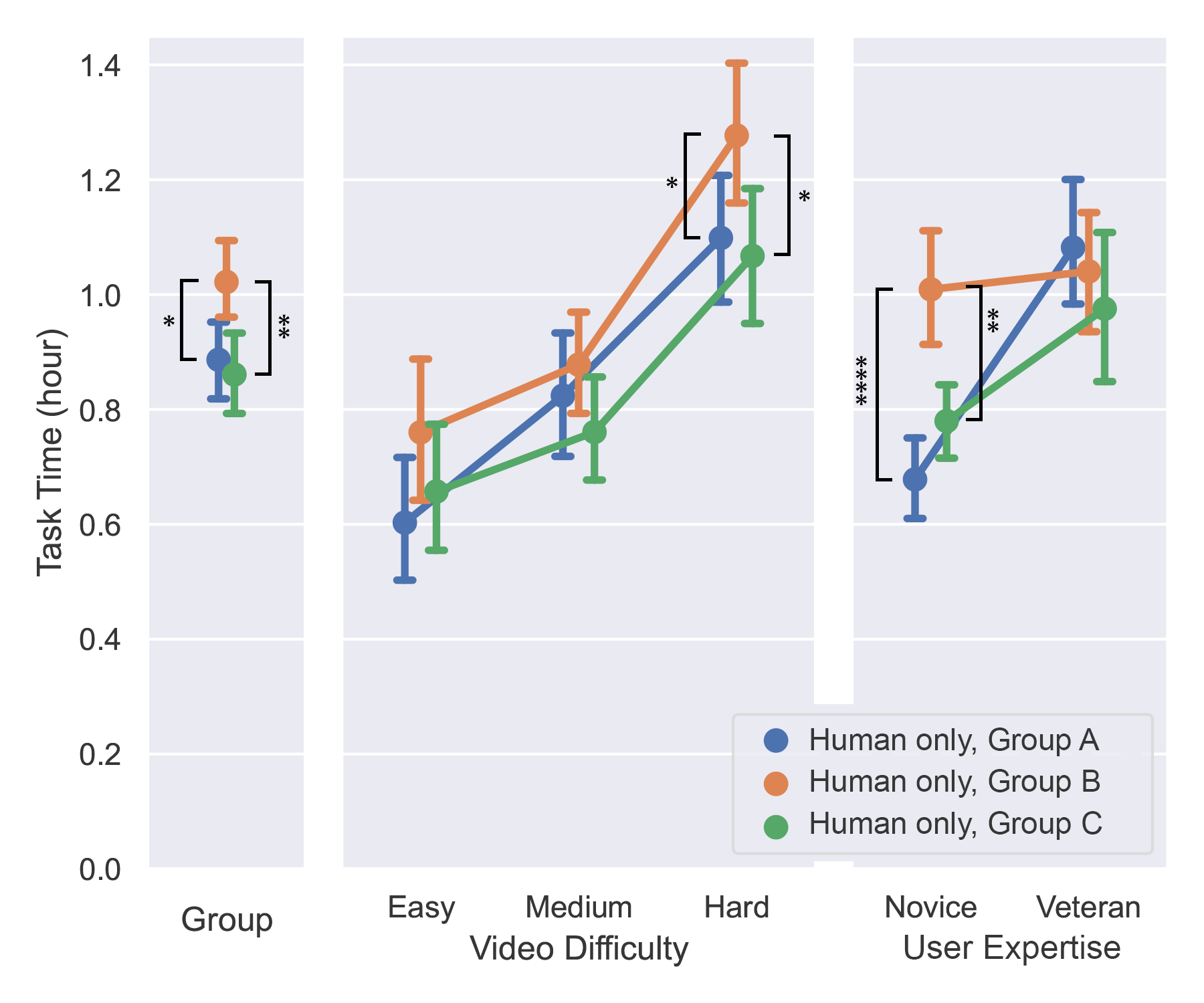}
    \vspace{-6mm}
    \captionof{figure}{Average annotation time for a single video in \textbf{Part~2}. After working 2-3 weeks on Part 1, every worker annotated another 12 videos in Part 2 but all used the same manual tool without AI recommendations. We no longer see a significant difference between Groups A~\&~C but Group B is now slower in hard videos, mainly caused by novice workers.}
    \label{fig:part2_task_time}
    \Description{Part 2 task completion time of the three methods, details described in the main text.}
  \end{minipage}
  \vspace{4mm}
  \newline
  \begin{minipage}[t]{1.0\textwidth}
    \includegraphics[width=\textwidth]{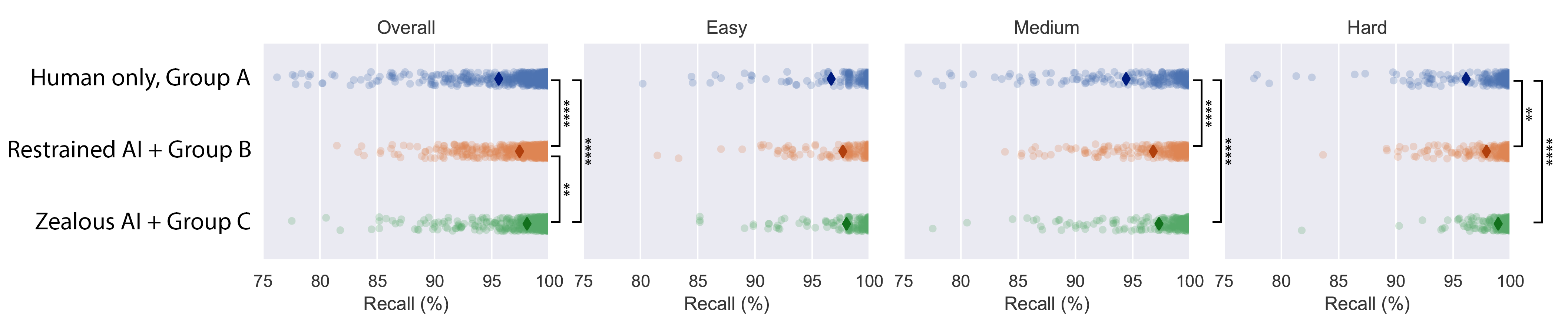}
    \vspace{-6mm}
    \captionof{figure}{The recall distribution of annotated videos in \textbf{Part 1}. For the purpose of visualization clarity, we plot the 75-100\% range in all recall distributions, which omits maximally 2\% of outlier cases. Higher recalls and a "shorter tail" are better. The average recall is marked with a darker diamond. The recall distribution reveals \textbf{the likelihood of having a higher quality result}, an insight needed to analyze results from crowdworkers. E.g., in hard videos (right), annotations from "zealous AI + Group C" have a shorter tail than other methods, as expected, the high-recall zealous AI recommendations make it easier for more people to achieve higher recalls especially when people's attention are pushed to the limit when there are three or more faces to track across many frames simultaneously.}
    \label{fig:part1_judgement_recalls}
    \Description{Part 1 recall distribution of the three methods, details described in the main text.}
  \end{minipage}
  \newline
  \begin{minipage}[t]{1.0\textwidth}
\hfill\includegraphics[width=1.0\textwidth]{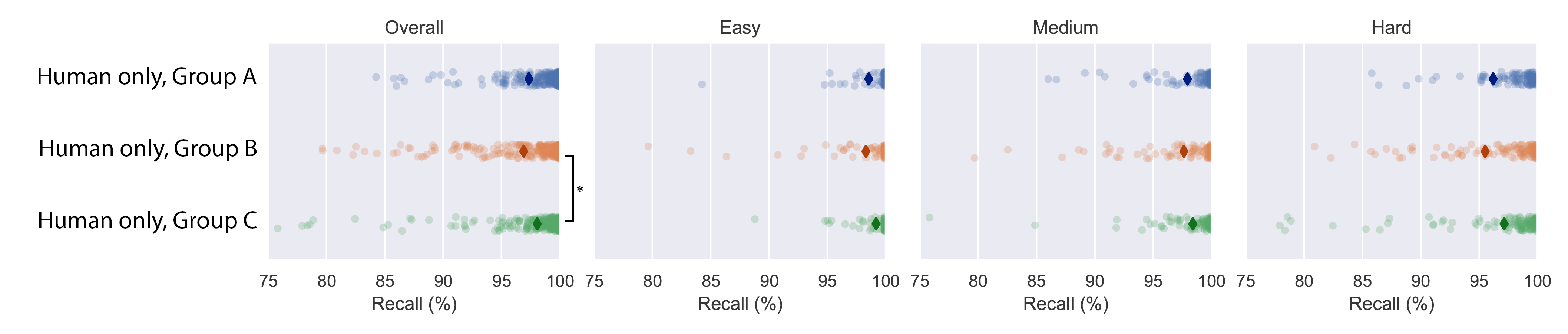}
    \vspace{-6mm}
    \captionof{figure}{The recall distribution of annotated videos in \textbf{Part 2}. The previously human-AI collaborative Groups B~\&~C no longer have access to the AI recommendations so they used the same manual method that Group A have been using. The overall subplot (left) shows visible longer tails from these two groups, especially Group C in hard videos (right), indicating a discrepancy in individuals' performance now without the help from AIs.}
    \label{fig:part2_judgement_recalls}
    \Description{Part 2 recall distribution of the three methods, details described in the main text.}
  \end{minipage}
\end{figure*}

\begin{figure*}[t]
  \centering
  \begin{subfigure}[b]{0.48\textwidth}
      \centering
      \includegraphics[width=\textwidth]{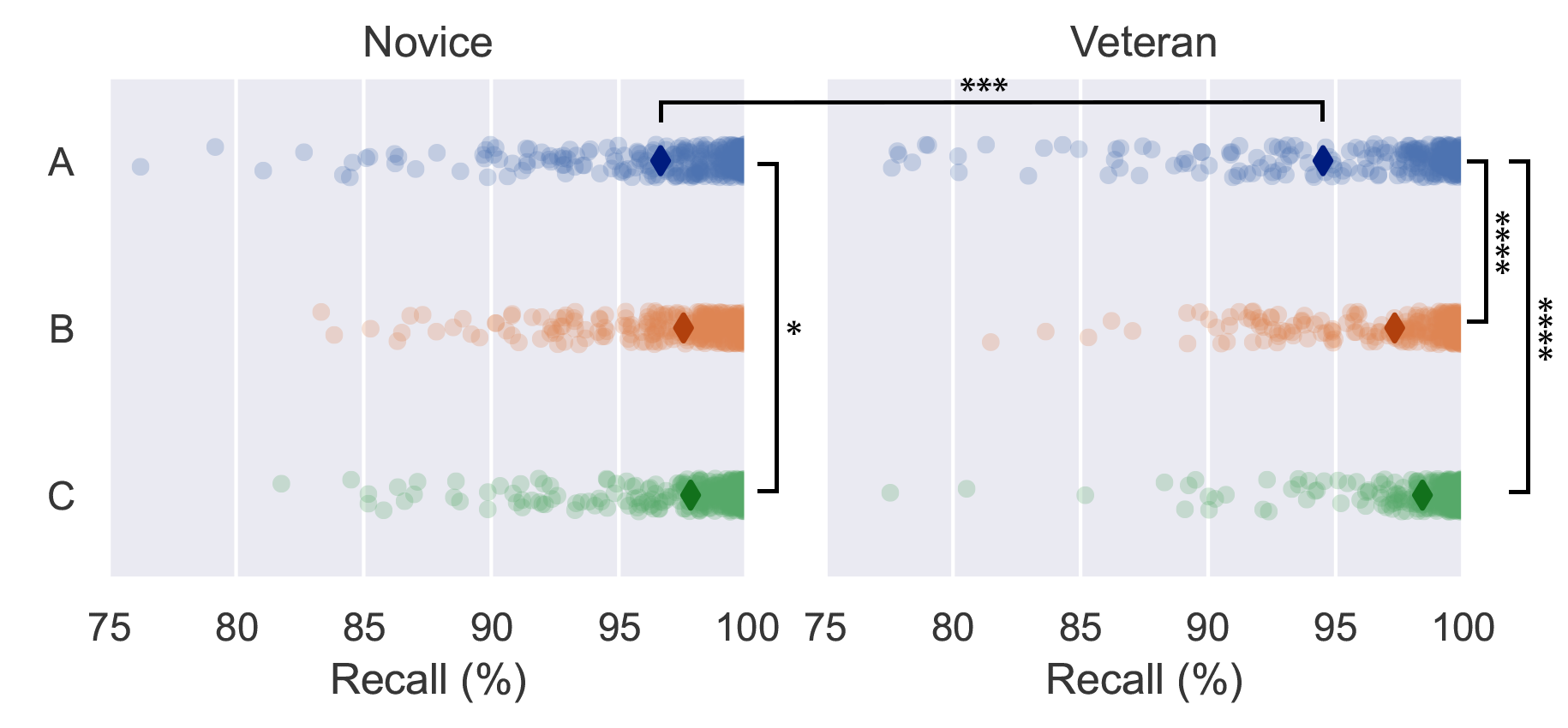}
      \caption{Part 1 (different methods between groups)}
  \end{subfigure}
  \quad
  \begin{subfigure}[b]{0.48\textwidth}  
      \centering 
      \includegraphics[width=\textwidth]{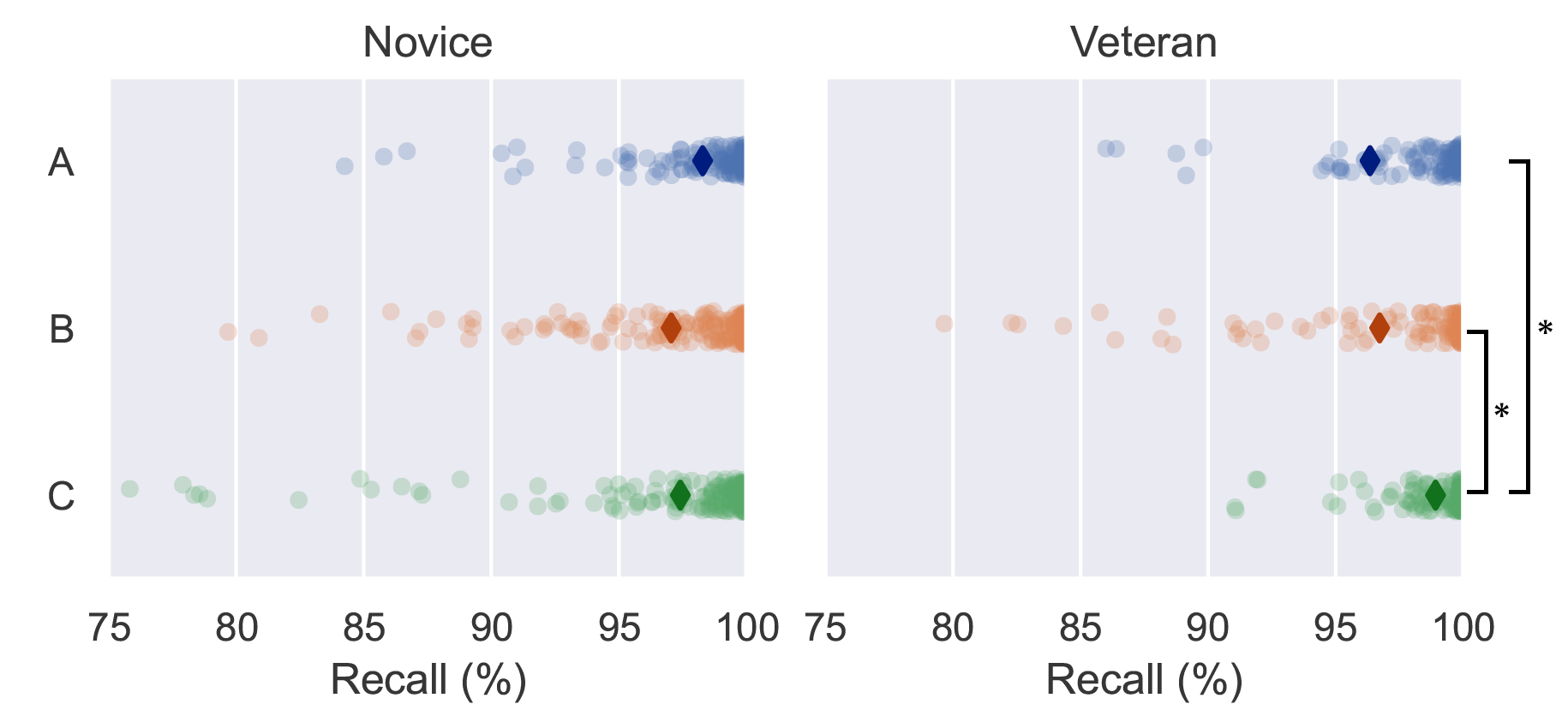}
      \caption{Part 2 (same method: human only)}
  \end{subfigure}
  \caption{\label{fig:veteran_recall}Recall distribution of annotated videos split by user expertise. Figure (a) shows both human-AI Groups~B~\&~C gained advantage over the manual method Group A mainly through veteran workers. The longer tails in Figure (b, novice) provide a new perspective to interpret Group C's long tails in Figure~\ref{fig:part2_judgement_recalls} (Overall) that the performance discrepancy is mostly caused by novice workers after they lost access to AI recommendations.}
  \Description{Part 1 and 2 recall distribution of annotated videos split by user expertise, details described in the main text.}
\end{figure*}

Research questions \textbf{Q1}, \textbf{Q2}, and \textbf{Q3} focus on results from Part 1 of the study (Figures~\ref{fig:team_effort}, \ref{fig:part1_task_time}, \ref{fig:part1_judgement_recalls}, and \ref{fig:veteran_recall}a), in which Groups B and C collaborated with restrained and zealous AIs. Question \textbf{Q4} focuses on results from Part 2 (Figures~\ref{fig:part2_task_time}, \ref{fig:part2_judgement_recalls}, and \ref{fig:veteran_recall}b) to examine how the prior human-AI collaboration experience could affect the users.


\subsection{Q1: Can the human-AI teams achieve "complementary team performance" in this task?}\label{complementary performance}

Bansal et al.\cite{Bansal_2021} defines \textit{complementary team performance} as the human-AI team performance exceeding both the human-only and AI-only performance. 

Figure~\ref{fig:team_effort} shows the two human-AI teams B \& C reached comparable F1 scores of 96.9\% \& 96.8\%,  respectively, significantly better than the human-only Group A that reached 94.5\% (Welch $F_{2, 1151}=18.2$, $p<0.0001$). Both human-AI teams improved F1 accuracy and recall significantly compared to their human-only counterpart. 

Because the high-stakes nature of this task rules out autonomous AI as a viable option, we really only need to compare the human-AI team performance with human-only performance in Part 1 of our study. However, to verify complementary team performance, we also verify that the two human-AI teams achieved higher performance in terms of F1 scores and recall than their respective AI's initial standalone performance.

Comparing each human-AI team with their perspective AI teammates' initial performance -- Group B annotators improved the restrained AI from 93.4\% to 96.9\% (Welch $F_{1, 1228}=178$, $p<0.0001$), Group C annotators improved the zealous AI from 90.9\% to 96.8\% (Welch $F_{1, 837}=169$, $p<0.0001$). Both human-AI teams improved significantly from their respective AI teammate's solo performance.

It is understandable that Bansal et al. only considered accuracy and did not compare task completion time in complementary performance, since the human-AI teamwork will undoubtedly add more time than AI alone. As we discussed, task completion times directly affect the operation cost as people are paid at an hourly rate, making it a critical metric for annotation tasks, so we additionally compare the human-AI teams' task completion times with the human-only team.

We saw overall significant differences between all three groups on task completion time (Welch $F_{2,1039}=48.6$, $p<0.0001$), as shown in Figure~\ref{fig:part1_task_time}, left. As a baseline, on average it took 1.05 hours for Group A to manually annotate a 30-second video of 900 frames. Group B took a significantly shorter time of 0.91 hours (Games-Howell $p<0.001$) to review the restrained AI recommendations. Group C only used 0.73 hours to review zealous AI's recommendations, also significantly shorter than the human-only Group A (Games-Howell $p<0.0001$). 

It is also worth noting that Group C, the zealous human-AI team, had an overall significantly worse starting point than Group B in terms of F1 score: 90.9\% vs. 93.4\% (Welch $F_{1, 854}=35.32$, $p<0.0001$) as shown in Figure~\ref{fig:team_effort}. However, annotators working with the zealous AI managed to achieve a significantly higher improvement in F1 score of +5.9\%~vs.~+3.5\% (Welch $F_{1, 934}=45.02$, $p<0.0001$) in significantly less time! This disadvantage for Group C provided the opportunity to demonstrate that our deduction in Section~\ref{sec:FPR} was correct -- a human-AI team can do better in both time and quality (in terms of F1 improvement) by asking the human to \textit{\textbf{reject}} more false positives and only \textit{\textbf{solve}} the most challenging faces, i.e., the high-recall zealous AI.

In summary, we have not only verified complementary team performance on accuracy, but also showed human-AI teams could achieve significantly shorter task completions time in a real-world case study.


\subsection{Q2: Which AI helps annotators be more efficient, i.e. save time?}

We mentioned that the professional \textbf{annotators are paid at their fixed hourly rate in this task}, which means 1)~they are not necessarily motivated to work faster, and 2) from the business perspective, their task completion time directly impacts operation costs. We discussed in Section~\ref{complementary performance} that overall, both human-AI teams have significantly shortened task completion time compared to the baseline Group A (Figure~\ref{fig:part1_task_time} left). Specifically, the zealous AI recommendations help annotators use 20\% less time than the restrained AI recommendations with statistical significance (0.73 hours vs. 0.91 hours, Games-Howell $p<0.0001$).

\subsubsection*{\textbf{Video difficulty.}} Figure~\ref{fig:part1_task_time} (middle) plots task time by video difficulty and saw a significant interaction between group and video difficulty on task completion time (ANOVA $F_{4, 1577}=5.37$, $p<0.0001$, $\eta_p^2=0.016$, small). Specifically, Group C which reviewed zealous AI recommendations used significantly less time than both Group A and B in medium videos (Bonferroni $p<0.0001$ \& $p<0.0001$), as well as in hard videos (Bonferroni $p<0.0001$ \& $p<0.01$). But no significant difference was found for easy videos among the three groups.

This observation matches very well with our expectations to different video difficulties: the built-in linear interpolation tool for manual annotation is very efficient in tracking a single face continuously, but \textbf{AI recommendations can dramatically reduce task time when tracking multiple faces simultaneously in medium and hard videos}. This finding allows the system designer to optimize efficiency further: if we know a certain portion of the data has one or fewer people in each frame, it would be reasonable to bypass the AI pre-annotation to save on the GPU budget.

\subsubsection*{\textbf{User expertise.}} When solely considering the user expertise factor, we were surprised that veteran workers are overall significantly slower than novice workers in both parts of the study (Welch, Part 1: $F_{1, 1380}=85.6, p<0.0001$, Part 2: $F_{1, 665}=22.2, p<0.0001$)! However, if we consider how people are paid, this result would be a reasonable optimization given the incentives -- veteran workers know the acceptable work pace, so they do not need to work faster than necessary. We further discussed worker's incentives in Section~\ref{sec:lower-bound}.


When we consider the group and user expertise factors at the same time, as shown in Figure~\ref{fig:part1_task_time} (right), both novice and veteran workers in Group C who reviewed the zealous AI recommendations were significantly faster than the baseline (Bonferroni $p<0.0001$ \& $p<0.0001$), while only the veterans in Group B finished faster (Bonferroni $p<0.0001$). This allows us to infer that, unlike the restrained AI that helps veterans more, \textbf{the zealous AI can consistently improve user completion time for both novice and veteran annotators}.


\subsection{Q3: Which AI helps annotators achieve higher recall?}
\label{sec:Q3}

From the F1 scores in Figure~\ref{fig:team_effort} we know that both AI-assisted methods yield significantly higher-quality annotations than the baseline method (compared in Section~\ref{complementary performance}), yet we saw no clear winner between the two human-AI teams. Because recall is paramount in video anonymization tasks, we analyze Group B and C's recall performance in detail.

Figure~\ref{fig:part1_judgement_recalls} shows that Group C, the annotators who reviewed zealous AI recommendations, have an overall significant advantage over Group B, which reviewed restrained AI recommendations (Games-Howell $p<0.01$). Interestingly, we noticed \textbf{a visible shorter tail} in Group C's recall distribution in hard videos (Figure \ref{fig:part1_judgement_recalls}, right). This observation matches the very nature of zealous AI -- giving more recommendations, even low-confidence ones, so the human teammate is less likely to miss a face. This strategy is especially effective in hard videos because tracking too many faces simultaneously pushes the user's attention to its limit. \textbf{Zealous AI's superfluous recommendations allow the user to focus on the action of \textit{reject}, rather than searching for missing faces and then \textit{solve}}.


Taking user expertise into account, Figure~\ref{fig:veteran_recall} (a) reveals that while both AIs improved the veterans' recall performance compared to the baseline Group A (Bonferroni A/B: $p<0.0001$, A/C $p<0.0001$), for novice workers, we only saw a significant advantage of Group C over Group A (Bonferroni $p<0.048$). It corroborates our previous finding on completion time that "the zealous AI can consistently improve both novice and veteran annotators" and extends the statement to higher recalls percentages as well.


\subsection{Q4: Will collaborating with an AI improve or hurt user skills?}
\label{sec:Q4}


Should the annotators lose access to their AI teammates in the future, how will they perform? While we are interested in improving human-AI team performance, we should also seriously consider how the prior human-AI collaboration experience would affect people's skills in the long run before deploying a new system.



To find out, we removed AI recommendations from Groups B and C in Part 2, so all groups now work with the manual tool that they have always been using for other projects. It took most annotators two to three weeks to complete Part 1 of the study. For the sake of interpreting the results of Part 2, we can consider this period a training period and their performance in Part 2 showcasing the effect of this medium-term training effort. 



Both Groups B~\&~C collaborated with their perspective AI teammates for 2-3 weeks, \textbf{but the restrained-AI-trained annotators in Group B performed worse than their peers in different ways} -- the novice workers were significantly slower than both A~\&~C, especially in hard videos. The veteran workers' annotations had lower recall percentages than the zealous-AI-trained workers in Group C.


\subsubsection*{\textbf{Completion time.}} Figure~\ref{fig:part2_task_time} shows the task completion time of Part 2's 12 new videos without AI recommendations. In all video difficulties, Group~C, annotators who previously worked with the zealous AI in Part~1, managed to finish as quickly as Group A, the annotators who were trained using the very manual method now in deployment for all groups. It shows that training with zealous AI recommendations does not negatively affect users' task completion time on subsequent manual tasks.

However, we were surprised to see that Group B annotators trained with the restrained AI became overall significantly slower than Groups A \& C (Tukey-HSD A/B: $p~<0.021$, B/C: $p<0.01$), and more specifically in hard videos (Bonferroni A/B: $p<0.044$, B/C: $p<0.013$). Figure~\ref{fig:part2_task_time} (right) shows that the effects stem mainly from the novice users (Bonferroni A/B: $p<0.0001$, B/C: $p<0.01$).

\subsubsection*{\textbf{Recall.}} On annotation quality, Figure~\ref{fig:part2_judgement_recalls} shows the Groups B annotators, trained by the high-precision restrained AI now produce lower-recall annotations (Games-Howell $p<0.05$) than Group C which was trained with the high-recall zealous AI. The user expertise breakdown shows the effect mostly comes from the veteran workers (Bonferroni $p<0.028$).

\subsubsection*{\textbf{What caused the negative training effect from the restrained AI?}} We would think that annotators in Group B should perform better in Part 2 of the study now that they have to manually annotate -- they practiced more on manually adding missing faces (\textit{\textbf{solve}}) working with the restrained AI recommendations. In contrast, Group C which trained with the zealous AI focused on \textbf{\textit{rejects}}. However, the experiment results show otherwise. Why was only Group B negatively affected? We believe there are two main factors in play:

\textbf{1) Not optimizing the AI teammate for the human-in-the-loop workflow.} Despite the fact that both AIs used the same face-detection model to generate the untracked bounding boxes in each frame for the tracker to process, the restrained AI recommendations were produced by ByteTrack \cite{zhang2022bytetrack} which is designed for autonomous tasks rather than for human-AI collaboration. We observed various issues using that tracker directly in pilot studies, so we proposed the FPR tracker specifically for a human-in-the-loop workflow with many optimizations with human users in mind (discussed in Section~\ref{sec:FPR}). Given the fact that only novice users became much slower in Part 2 of our study while veterans, who are more familiar with the annotation tool, were unaffected, we strongly believe that the negative transfer effect can be linked back directly to training with the restrained AI.

\textbf{2) Not optimizing the AI teammate for the task.} Recommendations from the high-precision restrained AI are naturally lower in recall than the zealous AI, i.e., the restrained AI missed more faces. Users who worked with such an AI for 2-3 weeks might actually have gotten used to the AI's pre-annotated videos (in Part 1) as "acceptable quality", thus matching their annotation effort with the less optimal recall when working on their own in Part 2. On the other hand, the zealous AI recommendations -- the high-recall AI more exhaustively demonstrated all faces that should be annotated, potentially raising the quality standard for the task.

In conclusion, various pieces of evidence from Part 2 of our study showed that despite decent human-AI team performance when working with the AI, naively deploying an AI system into a human-AI setting without considering the nature of the task or without optimizing it for the human teammates could lead to negative effects and potential deskilling of the users.

\section{Discussion}

\subsection{The key to forming a strong human-AI team}

We propose the restrained AI and the zealous AI to depict the tradeoff between precision and recall as two characteristics that have the potential of becoming advantages in human-AI teams if used properly. By actually using the annotation tools and watching annotators' screens for many hours, we observed that annotators need much less effort in improving precision than recall in a model-assisted annotation task, i.e., rejecting an incorrect box is much easier than adding a missing box, thus we should delegate more effort in improving recall to the AI so human only handles the most difficult boxes that the AI missed (Figure~\ref{fig:teaser}c). 

We think \textbf{an important insight from this study is that it is worthwhile to identify the complementary strengths of both human and AI teammates through an in-depth analysis of the task at hand}. While our observations can improve real-world object detection and tracking annotation tasks, in which correcting false-positive errors are easier for human, another task with a higher cost in correcting such errors could lead to different or even opposite optimizations. Working closely with end users can inspire us to decompose the AI's different properties (in our case precision and recall) and turn them into advantages to complement human skills. We hope this study can motivate fellow researchers to rethink existing AI assistance designs or at least the design for other video annotation tasks.

\begin{figure*}
  \noindent \begin{minipage}{0.6\textwidth}
    \includegraphics[width=\textwidth]{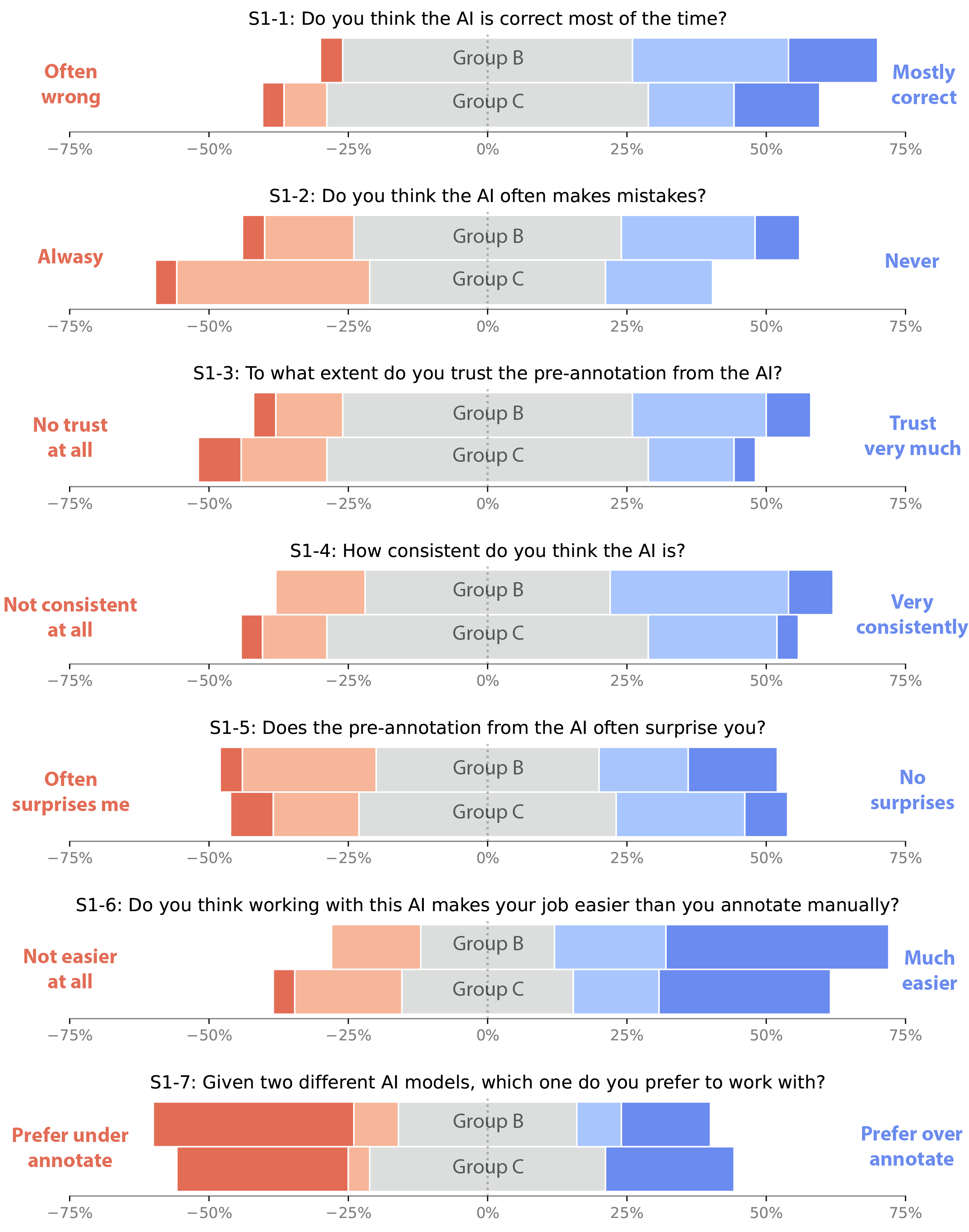}
    \captionof{figure}{Survey 1 (post-Part 1). We normalize each group's five-point Likert scale responses to 100\%. 0\% indicates no preference. In Part 1's between-subject study, annotators from Groups B \& C only worked with a single AI they were assigned to, so we do not compare the responses between B with C.}
    \label{fig:part1_survey}
    \Description{Survey 1 questions and responses.}
    \vspace*{25mm}
    \includegraphics[width=\textwidth]{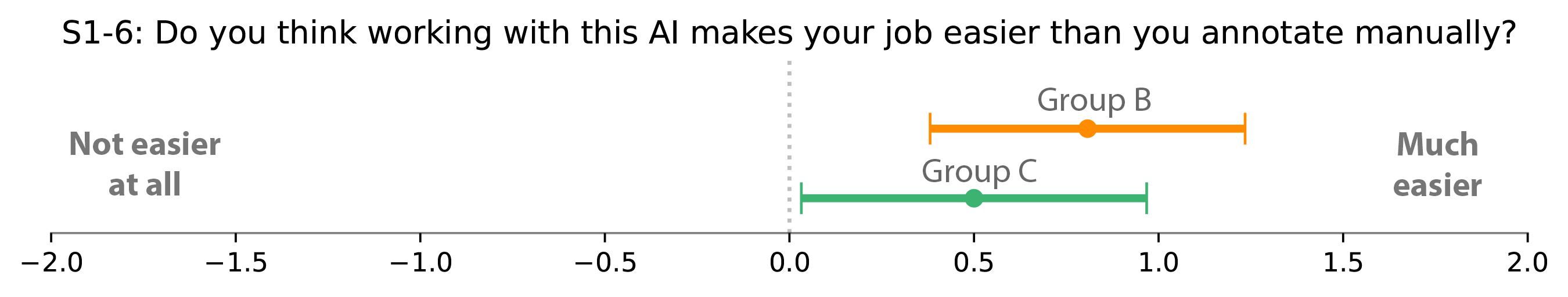}
    \captionof{figure}{Question S1-6 in Survey 1 indicates significant result. The five-point Likert scale responses are converted to [-2, 2] with mean and 95\% CI plotted.}
    \label{fig:part1_survey_q6}
    \Description{Survey 1 question 6 shows both group B and C think the AI made their annotation task easier.}
  \end{minipage}
  \hspace{0.01\textwidth}
  \begin{minipage}{0.38\textwidth}
    \includegraphics[width=\textwidth]{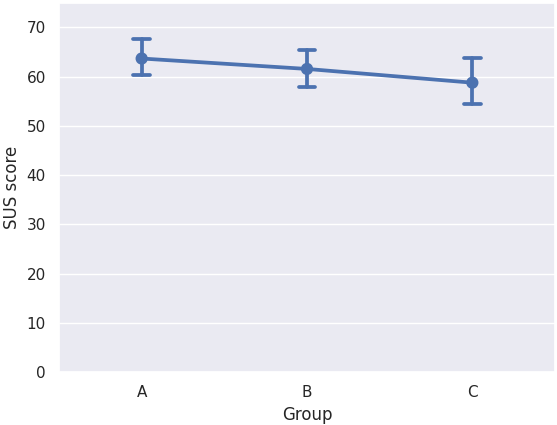}
    \captionof{figure}{A System Usability Scale (SUS) survey was administered at the conclusion of Part 1 of the study. But we saw  no significant difference between the groups. Similar to Survey 1 in Figure~\ref{fig:part1_survey}, participants tend to provide neutral feedback.}
    \label{fig:part1_sus}
    \Description{SUS score of the three methods.}
    \vspace*{53mm}
    \includegraphics[width=\textwidth]{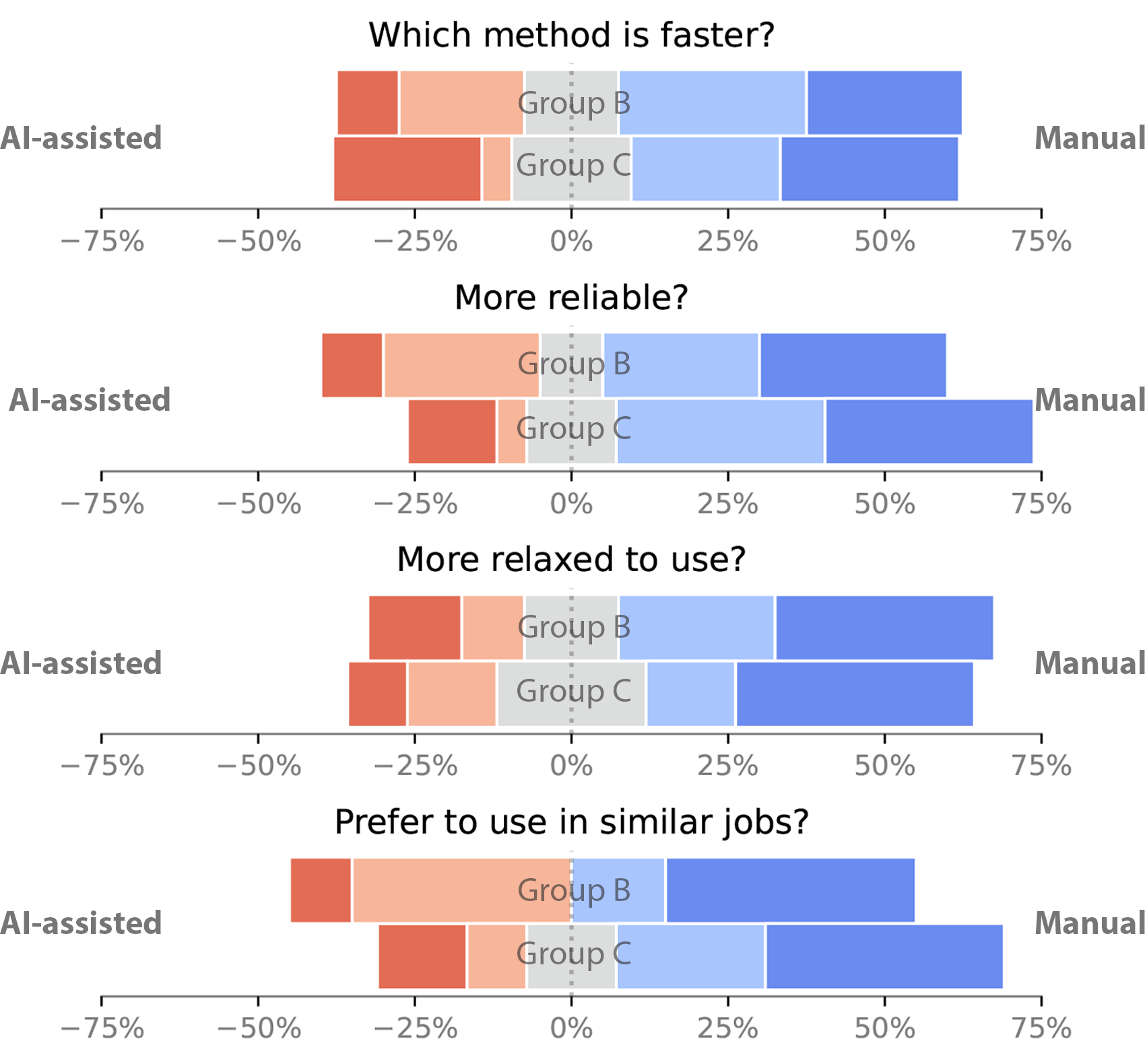}
    \captionof{figure}{Survey 2. Unlike Survey 1 in which annotators answered questions without comparison, Groups B \& C have used both AI-assisted and Manual methods at the end of Part 2. Thus this part of the study is close to a within-subject design where the independent variables are the AI-assisted and Manual method.}
    \label{fig:part2_survey}
    \Description{Survey 2 questions and responses.}
  \end{minipage}
\end{figure*}

\subsection{Can AI teammates set the quality lower bound in a crowdsourcing setting?}\label{sec:lower-bound}

We identified and rejected a single veteran user who submitted the majority of the low-quality annotations. This is an unexpected yet not surprising finding in a crowdsourcing setting: when paid at a flat hourly rate, people are not necessarily motivated to work faster. When lacking a quality-based performance evaluation mechanism, people are not necessarily motivated to push for "better-than-sufficient" quality.

However, could there be other users not making an effort in Groups B or C as well but not being identified? Because the two AIs have pre-annotated the videos in decent quality ($F1>90\%$), it's hard to tell if someone is actually happy with the AI's recommendations or is not pushing for even better quality.

What we know for sure is that such low-quality submissions, intentional or unintentional, will certainly appear in other real-world crowdsourcing tasks. However, in absence of ground truth, we won't be able to identify them in a real-world setting. It is also very costly to identify bad submissions -- ImageNet asks 10 votes for each image \cite{deng_imagenet_2009}, and Microsoft COCO asks 3-5 workers to judge each segmentation \cite{lin_microsoft_2014}.

Could the AI recommendations have played a critical role in preventing low-quality submissions, i.e., setting a lower bound for the annotation quality? While not verified in our study, this observation could provide yet another strong motivation for human-AI collaboration in a crowdsourcing setting. We encourage fellow researchers to consider this in future experiment designs.

\subsection{Seemingly contradictory survey results}
\label{sec:unexpected}


Figure~\ref{fig:part1_survey} shows user responses to the Survey 1 questions, with each group's five-point Likert scale responses normalized to 100\%. 0\% indicates no preference. Specifically, question \textit{S1-6} (Figure~\ref{fig:part1_survey_q6}) indicates that users from both human-AI teams, B and C, think that working with the AI makes the task easier than annotating manually. However, in Survey 2 (Figure~\ref{fig:part2_survey}), after users have tried both the AI-assisted and the Manual methods on the same task of similar videos, they express higher preference towards the Manual method regarding multiple aspects. As users took each survey immediately after Part 1 and Part 2 respectively, they might prefer the method they just used, but these responses from Groups B \& C are in conflict with their continued higher recall in Part 2.

Comparing Figure~\ref{fig:part1_judgement_recalls} (left) with Figure~\ref{fig:part2_judgement_recalls} (left), we observe that the Group B \& C annotators who had shorter tails in recall distribution than Group A in Part 1 ended up with longer tails in Part 2 after they lost the AI's assistance. It shows that a fraction of low-performing users were apparently held at a higher standard by the AI recommendations, and when the AI teammate was gone, they returned to their preferred standard. 


This observation might help explain the higher performance with the AI-assisted method but higher user preference for the Manual method. It also reminds us to take users' incentives into account when designing user preference questions in empirical studies -- It is well-known that the most favorable method is not necessarily the best performing method. We administered the System Usability Scale (SUS) survey and saw a trend to support this point in Figure~\ref{fig:part1_sus}, but the results are not significant.


\subsection{Limitations and Future Work}

\subsubsection*{\textbf{What are the conditions for which our findings hold?}} This study investigated a single high-stakes task that met the two aforementioned conditions: 1) either recall or precision is far more important than the other, and 2) the complementary strengths of human and AI can be identified and the precision-recall tradeoff can be exploited to improve the important metric for the given task. We proposed and observed that delegating more recall effort to the zealous AI can significantly improve team performance, which was mainly motivated by our observation that \textit{\textbf{reject}} is much easier than \textit{\textbf{solve}} for humans in AI-assisted annotation. Will our findings still hold if \textit{\textbf{reject}} is easier than \textit{\textbf{solve}} in a different task? What about precision-demanding tasks? We would love to see more HCI and AI researchers conduct latitudinal studies in multiple recall- or precision-demanding tasks to test and refine our findings.


\subsubsection*{\textbf{Tasks without high-performance models.}} Face detection is a well-studied problem with high-performance AI models. While we showed in Figure~\ref{fig:team_effort} that the AI and human can reach similar performance in this task to achieve complementary team performance, will our findings stand if either the human's performance or the AI's recommendations are much worse than the other? What is the lower bound F1 score limit for either the human or the AI to maintain complementary team performance? What are the F1 or precision/recall conditions for other researchers to reproduce our findings?

\subsubsection*{\textbf{Limitation from data and participants.}} We used a subset of realistic, egocentric video dataset \cite{Ego4D2022CVPR} in this study to measure with the skill of locating faces -- a human instinct that comes with relatively small inter-personal differences. However, could our findings still play a major role if the task was to identify and track other objects that could have larger inter-personal differences? Furthermore, working with amateurs via crowdsourcing platforms would introduce larger variances between individuals than with the professional workers employed in this study. Researchers would need to put more effort into benchmarking or measuring the human factor in such follow-up studies.




\subsubsection*{\textbf{Incentives for users to actively perform better.}} We discussed in Section~\ref{sec:unexpected}  observations that methods with better performances are not necessarily favored by the users. I.e., the users were involuntarily pushed to have higher performance by their AI teammates. From a system designer's perspective, the AI teammate should help users to voluntarily perform better given the right incentives.

\section{Conclusion}

In this work, we look beyond the accuracy of AI recommendations to explore a new direction to improve human-AI team performance -- the tradeoff between precision and recall in model tuning. We propose the concept of restrained and zealous AIs for high-precision and high-recall recommendations and conduct an experiment with 78 professional annotators to compare if and how the different AI recommendations can affect team performance in high-stakes human-AI collaboration. This work serves as a new example of complementary team performance in a large-scale realistic setting.

An in-depth analysis of the task helped us identify an optimization opportunity to harness complementary human and AI strengths utilizing the tradeoff between precision and recall in the AI model tuning -- given the importance of recall in face anonymization and the higher cost for humans to improve the recall in video annotation. We showed that the proposed high-recall zealous AI helps annotators achieve significantly better performance than the high-precision restrained AI in the video annotation task. Our follow-up study removed AI assistance and observed potentially negative training effects to the users -- if an AI is naively paired with humans without optimizing it for the task at hand or for the human-AI workflow. We feel these findings have important implications for the design of AI assistance in recall-demanding scenarios. We hope this work can also inspire researchers to look for additional directions in model tuning to improve human-AI team performance.

\begin{acks}
This work was partially done during the first author’s research internship at Appen. We thank all anonymous reviewers for their insightful comments and suggestions. We thank Huan Liu for her support and hand-drawn figures, Yue He and Yuedong Wang for their time and discussion, and members of the UCSB Four Eyes and Expressive Computation Laboratories for their helpful feedback. This work was partially supported by ONR awards N00014-19-1-2553 and N00014-23-1-2118.
\end{acks}

\bibliographystyle{ACM-Reference-Format}
\bibliography{references}

\appendix

\end{document}